\documentclass[doublecol]{epl2}

\title{Tsallis uncertainty}
\shorttitle{Tsallis uncertainty} 

\author{H. Moradpour$^1$\footnote{hn.moradpour@maragheh.ac.ir}, A. H. Ziaie$^1$\footnote{ah.ziaie@maragheh.ac.ir} \and C. Corda$^2$\footnote{cordac.galilei@gmail.com}}
\shortauthor{H. Moradpour, C. Corda and A. H. Ziaei}

\institute{$^1$ Research Institute for Astronomy and Astrophysics
of Maragha (RIAAM), University of Maragheh, P.O. Box 55136-553,
Maragheh, Iran\\
$^2$ International Institute for Applicable Mathematics and Information Sciences, B. M. Birla Science Centre, Adarshnagar, Hyderabad 500063 (India) and Istituto Livi, Via Antonio Marini, 9, 59100
Prato (Italy)}

\abstract{It has been recently shown that the Bekenstein entropy
bound is not respected by the systems satisfying modified forms of
Heisenberg uncertainty principle (HUP) including the generalized
and extended uncertainty principles, or even their combinations.
On the other, the use of generalized entropies, which differ from
Bekenstein entropy, in describing gravity and related topics
signals us to different equipartition expressions compared to
usual one. In that way, The mathematical form of an equipartition theorem can be related to the algebraic expression of a particular entropy, different from the standard Bekenstein entropy, initially chosen to describe the black hole event horizon, see E. M. C. Abreu et al.,  MPLA 32, 2050266 (2020).
Motivated by these works, we address three new uncertainty
principles leading to recently introduced generalized entropies.
In addition, the corresponding energy-time uncertainty relations
and Unruh temperatures are also calculated. As a result, it seems
that systems described by generalized entropies, such as those of
Tsallis, do not necessarily meet HUP and may satisfy modified
forms of HUP.}

\begin{document}

\maketitle

\section{Introduction}

Black holes are today considered theoretical laboratories where researchers can perform quantum gravity analyses. It is indeed a general conviction that, in some respects, black holes play the same role in gravitation that atoms played in the nascent quantum mechanics \cite{Bekenstein}.  The famous approach of Bekenstein \cite{Bekenstein 1}  and Hawking \cite{Hawking}  on black holes physics  also found a remarkable connection between gravity and thermodynamics. Fundamental thermodynamic quantities like entropy and temperature can be indeed associated with the black hole horizon. In this framework, Bekenstein also found an upper limit on the entropy  which can be contained within a given finite region of space having a finite amount of energy \cite{Bekenstein 2}. This is the so-called Bekenstein entropy bound, which implies that the information  necessary to describe a physical system, must be finite if the region of space and the energy is finite. The connection between gravity and thermodynamics  had an intriguing confirmation when Jacobson \cite{Jacobson} derived the Einstein field equation of general relativity by assuming that the Bekenstein bound and the laws of thermodynamics are correct.
On the other hand,  the Bekenstein entropy bound could not be respected by  systems satisfying modified HUP which include the generalized and extended uncertainty principles, or even their combinations \cite{Bek,maju1,ahmed1,ahmed2,ahmed3,taw,prdgup,eup1,eup2,eup3}.
In fact, the use of generalized entropies, which differ from Bekenstein entropy, in describing gravity and related topics show different equipartition expressions compared to the usual one. The mathematical form of an equipartition theorem may be related to the algebraic expression of a particular entropy, different from the standard Bekenstein entropy, initially chosen to describe the black hole event horizon \cite{sm1}.
In this letter, three new uncertainty principles leading to recently introduced generalized entropies are analysed. The corresponding energy-time uncertainty relations and Unruh temperatures are also calculated. Hence, one argues that systems described by generalized entropies, such as those of
Tsallis, do not necessarily meet HUP and may satisfy modified forms of HUP.

\section{Gaussian Extended Uncertainty Principle (GEUP)}

Quantum mechanics respects HUP expressed as

\begin{eqnarray}\label{1}
\Delta x\Delta p\geq\frac{1}{2},
\end{eqnarray}

\noindent leading to $\Delta p\geq\frac{1}{2\Delta x}$ combined
with $E\sim\Delta E\approx\Delta p$
\cite{ahmed1,ahmed2,ahmed3,deltax3,deltae2,deltae1,epl,eup2,taw},
$\Delta A\geq8\pi ER$ \cite{deltax1,deltax2,deltax3,taw}, and
$\Delta A_{min}\geq8\pi E \Delta x$
\cite{deltax2,deltax3,ahmed3,taw}, to reach

\begin{eqnarray}\label{2}
\Delta A_{min}\geq4\pi,
\end{eqnarray}

\noindent for $R\simeq\Delta x$
\cite{deltax1,deltax2,deltax3,taw,epl}, as the lower bound on
changes in the area of a black hole during the process of particle
absorption. Moreover, relying on relation $(\Delta
x)^2\approx\frac{A}{\pi}$
\cite{deltax1,deltax2,deltax3,ahmed3,taw}, we get $\Delta
A_{min}\simeq4\pi\lambda$ where $\lambda$ is unknown parameter \cite{ahmed3,taw}
that will be evaluated later.
Since during the process of particle absorption, the
entropy change of black hole equals to $\Delta S_{min}=b=\ln2$
\cite{ver,deltaamin,ahmed3}, one can easily write

\begin{eqnarray}\label{3}
\frac{dS}{dA}\equiv\frac{\Delta S_{min}}{\Delta
A_{min}}=\frac{b}{4\pi\lambda}\Rightarrow
S=\frac{bA}{4\pi\lambda},
\end{eqnarray}

\noindent apart from an integration constant, recovering the
well-known Bekenstein entropy ($S_{BE}=\frac{A}{4}$) if
$\lambda=\frac{b}{\pi}$, meaning that Bekenstein entropy bound is
consistent with HUP, and indeed, any changes in HUP modify this
entropy bound and affect the related physical outcomes
\cite{Bek,maju1,ahmed1,ahmed2,ahmed3,taw}.
More in general, $\lambda$ can be calculated from the black hole
mass spectrum. In fact, one recalls that the black hole horizon area
$A$ is related to the mass by the relation $A=16\pi M^{2}.$ If one
assumes that a neighboring particle is captured by the black hole
causing a transition from the state with $n$ to the state with $n+1$
(two neighboring levels). Thus, one immediately gets the area quantum
as
\begin{equation}
\triangle A_{n\rightarrow n+1}=16\pi\left(M_{n+1}^{2}-M_{n}^{2}\right).\label{eq: area spectrum}
\end{equation}
Let the mass eigenvalues $M_{n}$ be Bekenstein-like \cite{Bekenstein}
\begin{equation}
M_{n}=\gamma\sqrt{n},\label{eq: autovalori con moltiplicit=0000E0}
\end{equation}
where, in general, $\gamma\sim1$. Thus, we get
\begin{equation}
\triangle A_{n\rightarrow n+1}=16\pi\gamma^{2},
\end{equation}
that implies
\begin{equation}
4\pi\lambda=16\pi\gamma^{2}\Longrightarrow\lambda=4\gamma^{2}.\label{eq: lamda gamma}
\end{equation}
Bekenstein found \cite{Bekenstein}
\begin{equation}
M_{n}=\sqrt{\frac{n}{2}}\Longrightarrow\gamma^{2}=\frac{1}{2}\label{eq: Bekenstein}
\end{equation}
by using the Bohr-Sommerfeld quantization condition because he argued
that the black hole behaves as an adiabatic invariant. Thus, he obtained
$\triangle A_{n\rightarrow n+1}=8\pi$ \cite{Bekenstein}.

Now, let us consider extended uncertainty principle (EUP)
\cite{prdgup,eup1,eup2,eup3,Bek,taw}

\begin{eqnarray}\label{4}
\Delta x\Delta p\geq\frac{1}{2}[1+\frac{\beta\pi}{4}(\Delta x)^2],
\end{eqnarray}

\noindent giving the non-zero minimum uncertainty in momentum as
$(\Delta p)_{min}=\sqrt{\frac{\beta\pi}{4}}$ \cite{hassan1}. Here,
$\beta$ is a positive parameter \cite{prdgup,taw}, and the above
recipe leads to $\Delta A_{min}\simeq4\pi\lambda[1+\frac{\beta}{4}
A]$, and finally, the R\'{e}nyi entropy
$S=\frac{b}{\pi\lambda\beta}\ln[1+\frac{\beta}{4}A]$ for
$\lambda=\frac{b}{\pi}$ \cite{epl}. Thus, it seems that the
entropy of black hole depends on the uncertainty principle
relation and HUP is not, probably, allowed whenever Bekenstein
entropy is not respected by black hole.

Recently, working in the Tsallis statistics framework, a new
entropy has been obtained for black holes as follows
\cite{mah2019,me,kav}

\begin{eqnarray}\label{5}
&&S_q^T=\frac{1}{1-q}[\exp\big((1-q)S_{BE}\big)-1],
\end{eqnarray}

\noindent in which $q$ is the non-extensive parameter fixed by
using other parts of physics or experiments
\cite{pla,tsa,mah2019,me,kav}, and Bekenstein entropy is recovered
at the appropriate limit of $q\rightarrow1$, where the Gibbs
statistics works \cite{me}. This entropy has theoretically proper
power to model the cosmic history in various setups \cite{me,kav}.

Now, relying on the above recipe, one can easily see that, apart
from an integration constant, Eq.~(\ref{5}) can be obtained from a
gaussian generalization of the extended uncertainty principle as

\begin{eqnarray}\label{6}
\Delta x\Delta p\geq\frac{1}{2}\exp[-\frac{(\Delta
x)^2}{\sigma^2}]=\frac{1}{2}(1-\frac{(\Delta x)^2}{\sigma^2}+..)
\end{eqnarray}

\noindent where $\sigma\equiv\sqrt{\frac{1}{\delta\pi}}$,
$\delta\equiv1-q$, and clearly, HUP is recovered whenever
$\delta\rightarrow0$. Additionally, the above GEUP leads to a
minimum in momentum uncertainty as

\begin{eqnarray}\label{7}
(\Delta
p)^{min}=\sqrt{\frac{e}{-2\sigma^2}}=\sqrt{\frac{-\delta\pi
e}{2}},
\end{eqnarray}

\noindent indicating that the existence of non-extensivity
parameter $q$ admits a non-zero value for $(\Delta p)^{min}$.
Indeed, by comparing this result and the last equality of
Eq.~(\ref{6}) with $(\Delta p)_{min}$ and Eq.~(\ref{4}),
respectively, one finds $(\Delta
p)^{min}=\sqrt{\frac{e}{2}}(\Delta p)_{min}$ whenever
$\beta=-4\delta$. \textit{It is also worthwhile mentioning that
some implications of GEUP on quantum mechanics and statistics of
ideal gases are studied in Ref.~\cite{epjp}, and hence, one can
conclude that gravitational systems which meet the algebra
introduced in Ref.~\cite{epjp} obey~(\ref{5}) instead of
Bekenstein bound.} Therefore, it seems that there is a deep
connection between the quantum features of gravity and deviation
from Gibbs statistics, a result in agreement with previous reports
\cite{epl,me,mah2019,bar}.

\subsection{Gaussian Energy-Time Uncertainty (GETU) relation}

The exact form of energy-time uncertainty relation is still
controversial \cite{revet}, and it has also been shown that EUP
affects it \cite{amjp}. One key point in the above calculations is
$\Delta E\approx\Delta p$
\cite{ahmed1,ahmed2,ahmed3,deltax3,deltae2,deltae1,epl,eup2,taw},
an assumption reinforced by the fact that particles' velocity is so
close to that of light helping us in assuming $\Delta x\approx
\Delta t$ \cite{hassan1,amjp}. Inserting all of these in
Eq.~(\ref{6}), one can easily get GETU as

\begin{eqnarray}\label{8}
\Delta t\Delta E\geq\frac{1}{2}\exp[-\frac{(\Delta
t)^2}{\sigma^2}].
\end{eqnarray}

\subsection{Unruh temperature}

Assuming dispersion relation $E=p$ \cite{scar}, Eq.~(\ref{6})
leads to

\begin{eqnarray}\label{9}
\Delta x\Delta E\geq\frac{1}{2}\exp[-\frac{(\Delta
x)^2}{\sigma^2}].
\end{eqnarray}

\noindent Moreover, $E=ma\Delta x$ is the kinetic energy of a
particle with acceleration $a$ and mass $m$ that experiences
displacement $\Delta x$, and if it is assumed as the required
energy for creating $N$ pairs of the primary considered particle
from vacuum, then we have \cite{scar}

\begin{eqnarray}\label{10}
2N\simeq a\Delta x,
\end{eqnarray}

\noindent combined with Eq.~(\ref{9}) and $\Delta
E\simeq\frac{3}{2}T$ \cite{scar}, to reach

\begin{eqnarray}\label{11}
T=T_U\exp[-\frac{1}{(3T_U\sigma)^2}]=T_U-\frac{1}{9T_U\sigma^2}+....,
\end{eqnarray}

\noindent as the modified Unruh temperature. In order to obtain
this result, we used the fact that Unruh temperature
$T_U(\equiv\frac{a}{2\pi})$ should be obtained at the limit of
$q\rightarrow1$ helping us in finding $N=\frac{\pi}{3}$ in full
agreement with previous results \cite{scar}. If one started with
EUP instead of GEUP, then the first two terms of the last line of
Eq.~(\ref{11}) appeared with $\beta(=-4\delta)$, and hence, we got
the modified Unruh temperature in the presence of EUP
\cite{hassan1}. The effects of extended generalized uncertainty
principle expressed as \cite{eup1,hasan,prdgup}

\begin{eqnarray}
\Delta x\Delta p\geq\frac{1}{2}[1+\frac{\beta\pi}{4}(\Delta
x)^2+\gamma(\Delta p)^2],
\end{eqnarray}

\noindent on horizon entropy, Unruh temperature and energy-time
uncertainty relation have also been studied in Ref. \cite{hasan}.

\section{Power-law uncertainty principle (PUP)}

Using Tsallis statistics, another entropy is also obtainable for
black holes as \cite{tsallis}

\begin{eqnarray}\label{12}
S=\gamma A^\eta,
\end{eqnarray}

\noindent where $\gamma$ and $\eta$ are unknown free parameters.
Recently, Barrow found out that a black hole whose horizon has
multi-fractal structure satisfies this entropy bound \cite{bar}.
It is easily checkable that entropy is positive if $\gamma>0$ and
moreover, the second law of thermodynamics ($\Delta S\geq0$) is
also satisfied whenever $\eta>0$. On the other hand, a dynamical
analysis on a holographic dark energy model built by
using~(\ref{12}) implies on $\eta>1$ \cite{ebrahimi}.

In line with previous section, this entropy is achieved from the
PUP

\begin{eqnarray}\label{13}
\Delta x\Delta p\geq\frac{(\Delta x)^{2-2\eta}}{2},
\end{eqnarray}

\noindent when $\gamma\equiv\frac{b}{4\pi\lambda\eta}$, and
clearly, Bekenstein entropy is recovered when $\eta=1$, and
$\lambda=\frac{b}{\pi}$. Additionally, accepting $\eta>1$
\cite{ebrahimi}, we can see $\Delta
x\rightarrow0(\infty)\Leftrightarrow\Delta p\rightarrow\infty(0)$,
the same as what we have in the HUP regime. In this manner,
calculations for the modified energy-time uncertainty relation and
Unruh temperature lead to

\begin{eqnarray}\label{14}
\Delta t\Delta E\geq\frac{(\Delta t)^{2-2\eta}}{2},
\end{eqnarray}

\noindent and

\begin{eqnarray}\label{15}
T=9^{\eta-1}T_U^{2\eta-1},
\end{eqnarray}

\noindent respectively. It is finally worthwhile mentioning that,
unlike the generalized uncertainty principle (GUP)-modified Unruh
temperature \cite{scar}, those obtained here are always real.

\section{Power-law extended uncertainty principle (PEUP)}

Accepting Bekenstein entropy as a Tsallis entropy \cite{1,2}, the
Sharma-Mittal entropy content of black hole is introduced as
\cite{sm}

\begin{eqnarray}\label{16}
S=\frac{1}{R}\big[(1+\frac{\delta
A}{4})^{\frac{R}{\delta}}-1\big],
\end{eqnarray}

\noindent in which $R$ is an unknown free parameter \cite{sm,sm1}.
Following the previous sections, PEUP and corresponding
energy-time and Unruh temperature are obtained as

\begin{eqnarray}\label{17}
\Delta x\Delta p\geq\frac{1}{2}\big(1+\delta\pi(\Delta
x)^2\big)^{1-\frac{R}{\delta}},
\end{eqnarray}

\noindent covering EUP when $R=0$,

\begin{eqnarray}\label{18}
\Delta t\Delta E\geq\frac{1}{2}\big(1+\delta\pi(\Delta
t)^2\big)^{1-\frac{R}{\delta}},
\end{eqnarray}

\noindent and

\begin{eqnarray}\label{19}
T=T_U\big(1+\frac{\delta\pi}{9T_U^2}\big)^{1-\frac{R}{\delta}},
\end{eqnarray}

\noindent respectively. Simple calculations also lead to

\begin{eqnarray}\label{20}
(\Delta
p)_{m}=\sqrt{\frac{\pi(\delta-R)}{4}}[\frac{2(\delta-R)}{\delta-2R}]^{\frac{\delta-R}{\delta}},
\end{eqnarray}

\noindent as the minimum momentum in this regime achievable for
$\Delta x=\sqrt{\frac{1}{\pi(\delta-2R)}}$. In agreement with the
EUP regime (i.e. $(\Delta p)_{min}$), Eq.~(\ref{20}) yields
$(\Delta p)_{m}=\sqrt{\pi\delta}$ whenever $R=0$.
\section{Final discussion and conclusion remarks}

Black holes have fundamental importance in quantum gravity. In fact, it is a general conviction, arising from an intuition of Bekenstein \cite{Bekenstein}, that they should play the same role in quantum gravity that atoms played in the nascent quantum mechanics.  Bekenstein \cite{Bekenstein 1}  and Hawking \cite{Hawking}  also found fundamental connections between gravity and thermodynamics in the framework of black hole physics. Remarkably, some of the most important thermodynamic quantities like entropy and temperature may be associated with the black hole horizon.  Bekenstein also found an upper limit on the entropy  which can be contained within a given finite region of space with a finite amount of energy \cite{Bekenstein 2}. The Bekenstein entropy bound implies that the information necessary to describe a physical system, must be finite in a finite region of space having finite energy. In 1995 Jacobson \cite{Jacobson} found an intriguing confirmation of the connection between gravity and thermodynamics  by deriving the field equations of general relativity through the  assumption that the Bekenstein bound and the laws of thermodynamics should be correct.

But the Bekenstein entropy bound could, in principle, not be respected if a system satisfied modified HUP including the generalized and/or extended uncertainty principles, or even their combinations \cite{Bek,maju1,ahmed1,ahmed2,ahmed3,taw,prdgup,eup1,eup2,eup3}.
Generalizations of Gibbs statistics indeed lead to different entropy-area relations compared with that of
Bekenstein proposal. Consequently, the use of generalized entropies, which differ from Bekenstein entropy, in the description of gravity and related topics, can show different equipartition expressions compared to the usual one. The mathematical expression of a generalized equipartition theorem may be related to the algebraic expression of a particular entropy, different from the standard Bekenstein entropy, initially chosen to describe the black hole event horizon \cite{sm1}. In general, the cited generalizations have significant and impressive predictions and outcomes in various branches of physics
such as cosmology, astrophysics, the physics of plasma and Sun, condensed matter physics and etc,. It has also previously been shown that if HUP is modified, then some physical properties and constraints of systems such as the Bekenstein entropy bound, energy-time uncertainty relation and Unruh temperature will also be modified.

In this letter, three new uncertainty principles leading to recently introduced generalized entropies have been analysed by also calculating the corresponding energy-time uncertainty relations and Unruh temperatures. Thus, it seems that systems described by generalized entropies, such as those of
Tsallis, do not necessarily meet HUP and may satisfy modified forms of HUP.  This is a confirmation that  modified versions of HUP lead to generalized entropies. Therefore, by employing previously introduced approaches in finding out modified:  $i$) energy-time uncertainty relations,  $ii$) Unruh temperatures, $iii$) horizon entropy by starting from EUP and GUP, one finds generalizations of HUP  that lead to some recently introduced generalized entropies. Implications of GEUP, GUP and PEUP on energy-time uncertainty relation, and also Unruh temperature have been studied which indicate that the modified
Unruh temperatures are always real and positive whenever GEUP and PUP are valid.

\section{Acknowledgements}
The Authors thank an unknown Referee for useful comments.

%
%
%
%
%

\begin{thebibliography}{99}
\bibitem{Bekenstein}J. D. Bekenstein, in Prodeedings of the Eight Marcel Grossmann Meeting, T. Piran and R. Ruffini, eds., pp. 92-111 (World Scientific Singapore 1999).
\bibitem{Bekenstein 1}J. D. Bekenstein, Lett. Nuovo Cimento, 11, 467 (1974).
\bibitem{Hawking}S. W. Hawking, Commun. Math. Phys. 43, 199 (1975).
\bibitem{Bekenstein 2}J. D. Bekenstein, Phys. Rev. D 23, 287 (1981).
\bibitem{Jacobson}T. Jacobson, Phys. Rev. Lett. 75, 1260 (1995).
\bibitem{ahmed3} A. Awada, A. F. Ali, JHEP 06, 093 (2014).
\bibitem{ahmed1} A. F. Ali, A. Tawfik, AHEP 2013, 126528 (2013).
\bibitem{ahmed2} A. F. Ali, Phys. Lett. B 732, 335 (2014).
\bibitem{taw} A. Tawfik, A. Diab, Int. J. Mod. Phys. D 23, 1430025 (2014).
\bibitem{deltae1} F. Scardigli, Nuovo Cimento B 110, 1029 (1995).
\bibitem{deltae2} R. J. Adler, P. Chen, D. I. Santiago, Gen. Relativ. Gravit. 33, 2101 (2001).
\bibitem{deltax3} M. Cavaglia, S. Das, Class. Quant. Grav. 21, 4511 (2004).
\bibitem{eup2} M. I. Park, Phys. Lett. B 659, 698 (2008).
\bibitem{epl} H. Moradpour et al., EPL. 127, 60006 (2019).
\bibitem{deltax1} A. J. M. Medved, E. C. Vagenas, Phys. Rev. D 70, 124021 (2004).
\bibitem{deltax2} B. Majumder, Phys. Lett. B 703, 402 (2011).
\bibitem{deltaamin} J. D. Bekenstein, Phys. Rev. D 7, 2333 (1973).
\bibitem{ver} E. P. Verlinde, JHEP 1104, 029 (2011).
\bibitem{Bek} F. J. Wang, Y. X. Gui, Y. Zhang, Gen. Relativ. Gravit. 41, 2381 (2009).
\bibitem{maju1} B. Majumder, Astrophys. Space. Sci. 336, 331 (2011).
\bibitem{prdgup} A. Kempf, G. Mangano, R. B. Mann, Phys. Rev. D 52, 1108 (1995).
\bibitem{eup1} B. Bolen, M. Cavaglia, Gen. Relativ. Gravit. 37, 1255 (2005).
\bibitem{eup3} C. Bambi, F. R. Urban, Class. Quantum. Grav. 25, 095006 (2008).
\bibitem{hassan1} W.~S.~Chung, H.~Hassanabadi, Phys. Lett. B \textbf{793}, 451 (2019)
\bibitem{mah2019} K. Mejrhit, S. E. Ennadifi, Phys. Lett. B 794, 24 (2019).
\bibitem{me} H. Moradpour, A. H. Ziaie, M. Kord Zangeneh, arXiv:2005.06271.
\bibitem{kav} K. Abbasi, S. Gharaati, arXiv:2006.01763
\bibitem{pla} M. Masi, Phys. Lett. A 338, 217 (2005).
\bibitem{tsa} C. Tsallis, J. Stat. Phys. 52, 479 (1988).
\bibitem{epjp} R. A. El-Nabulsi, Eur. Phys. J. Plus. 135, 34 (2020).
\bibitem{bar} J.~D.~Barrow, arXiv:2004.09444 [gr-qc].
\bibitem{revet} V. V. Dodonov, A. V. Dodonov, Phys. Scr. 90 074049 (2015).
\bibitem{amjp} J.~Denur, Am. J. Phys. 78, 1132 (2010).
\bibitem{hasan} H. Hassanabadi, N. Farahani, W. S. Chung, B. C. L\"{u}tf\"{u}o\u{g}lu, EPL 130, 40001 (2020).
\bibitem{scar} F.~Scardigli, M.~Blasone, G.~Luciano, R.~Casadio, Eur. Phys. J. C \textbf{78}, 728 (2018).
\bibitem{tsallis} C. Tsallis, L. J. L. Cirto, Eur. Phys. J. C 73, 2487 (2013).
\bibitem{ebrahimi} E. Ebrahimi, Astrophys. Space. Sci. 365, 92 (2020).
\bibitem{sm} A. Sayahian Jahromi et al., Phys. Lett. B 780, 21 (2018).
\bibitem{sm1} E. M. C. Abreu et al.,  MPLA 32, 2050266 (2020).
\bibitem{1} A. Majhi, Phys. Lett. B 775, 32 (2017).
\bibitem{2} N. Komatsu, Eur. Phys. J. C 77, 229 (2017).


\end{thebibliography}
\end{document}